\documentclass[aps,prb,reprint,superscriptaddress,longbibliography]{revtex4-2}

\usepackage{graphicx}
\usepackage{dcolumn}
\usepackage{bm}
\usepackage{xcolor}
\usepackage{hyperref}
\usepackage{amsmath}
\usepackage{soul}
\usepackage{lipsum}

\begin{document}

\preprint{APS/123-QED}

\title{Programmable transport of rotating particles in obstacle arrays}

\author{Marcos Puerto}
\affiliation{Department of Condensed Matter Physics, Universidad Autonoma de Madrid, 28049, Madrid, Spain}
\author{Alfredo Alexander-Katz}
\affiliation{Department of Materials Science and Engineering, Massachusetts Institute of Technology, Cambridge, MA, 02139, USA}
\author{Juan L. Aragones}
\email{juan.aragones@uam.es}
\affiliation{Department of Theoretical Condensed Matter Physics, Universidad Autonoma de Madrid, 28049, Madrid, Spain}
\affiliation{Condensed Matter Physics Center (IFIMAC), Universidad Autonoma de Madrid, 28049, Madrid, Spain}
\affiliation{Instituto Nicolas Cabrera, Universidad Autonoma de Madrid, 28049, Madrid, Spain}
\author{J.V. Alvarez}
\email{jv.alvarez@uam.es}
\affiliation{Department of Theoretical Condensed Matter Physics, Universidad Autonoma de Madrid, 28049, Madrid, Spain}
\affiliation{Condensed Matter Physics Center (IFIMAC), Universidad Autonoma de Madrid, 28049, Madrid, Spain}

\date{\today}

\begin{abstract}
Rotating colloids, or spinners, in obstacle arrays exhibit frequency-set stationary orbits and currents set by the competition between an inertial, Magnus-like lift and short-range attraction. Fully resolved lattice–Boltzmann simulations reveal the hydrodynamic coupling and identify the lift mechanism, while a symmetry-based Langevin model captures the resulting balance. In periodic lattices, the superposition of scalar and vector potentials produces two robust orbital regimes: corner states, in which spinners orbit individual posts, and inner states, in which orbits couple across four neighboring obstacles. Slow frequency modulation toggles these states and produces directed, stepwise transport across the grid. This establishes a minimal hydrodynamic mechanism, controlled by a single driving parameter, for programmable guidance of active rotors in structured environments.

\end{abstract}

\maketitle

\section{Introduction}
Active colloidal particles, from living organisms to synthetic beads, droplets or micro-robots, convert external or internal energy into motion. These self-driven objects can direct their motion in response to external stimuli such as chemical, mechanical or optical cues~\cite{berg1977,fortunato2024,mierke2020,lozano2016,koley}. When they traverse structured or crowded landscapes, the interplay between propulsion, hydrodynamic coupling and boundary interactions gives rise to transport modes with no analogue in equilibrium diffusion such as rectification~\cite{wanRectificationSwimmingBacteria2008,nikolaActiveParticlesSoft2016,Lauga2016}, directional locking~\cite{brun-cosme-brunyDeflectionPhototacticMicroswimmers2020,reichhardtDirectionalLockingEffects2020} or long-lived edge orbits~\cite{spagnolie2015a}.
Systems composed of rotating active particles, {\em spinners}, do not exhibit directed motion, but their activity generates azimuthal rotlet flows which significantly modify the dynamical properties of the system~\cite{Avron1998,Banerjee2017} and strongly influence its self-assembly~\cite{Grzybowski2002,Gotze2011,climentDynamicSelfAssemblySpinning2007,Goto2015,Aragones2019,fily2012b}. Near boundaries, these flows couple rotation and translation, enabling spinners to glide along walls or obstacles~\cite{gorceRollingSpinnersWater2021, Gotze2011}. In such systems, transport is determined by the environment structure, opening the door to the design of systems with programmed and/or optimized transport properties. Additionally, even in the overdamped realm of microns and milliseconds, weak but finite inertia can trigger lift forces that act on the microparticle~\cite{Rubinow1961,saffman1965,Goto2015,Aragones2016,caoMemoryinducedMagnusEffect2023}, which would push it away from the boundary, while electrostatic or depletion attractions can pull it back. This balance between non-dissipative effects provides dynamical control over the transport properties~\cite{Yazdi2020}.

\begin{figure}
\includegraphics[width=1\linewidth]{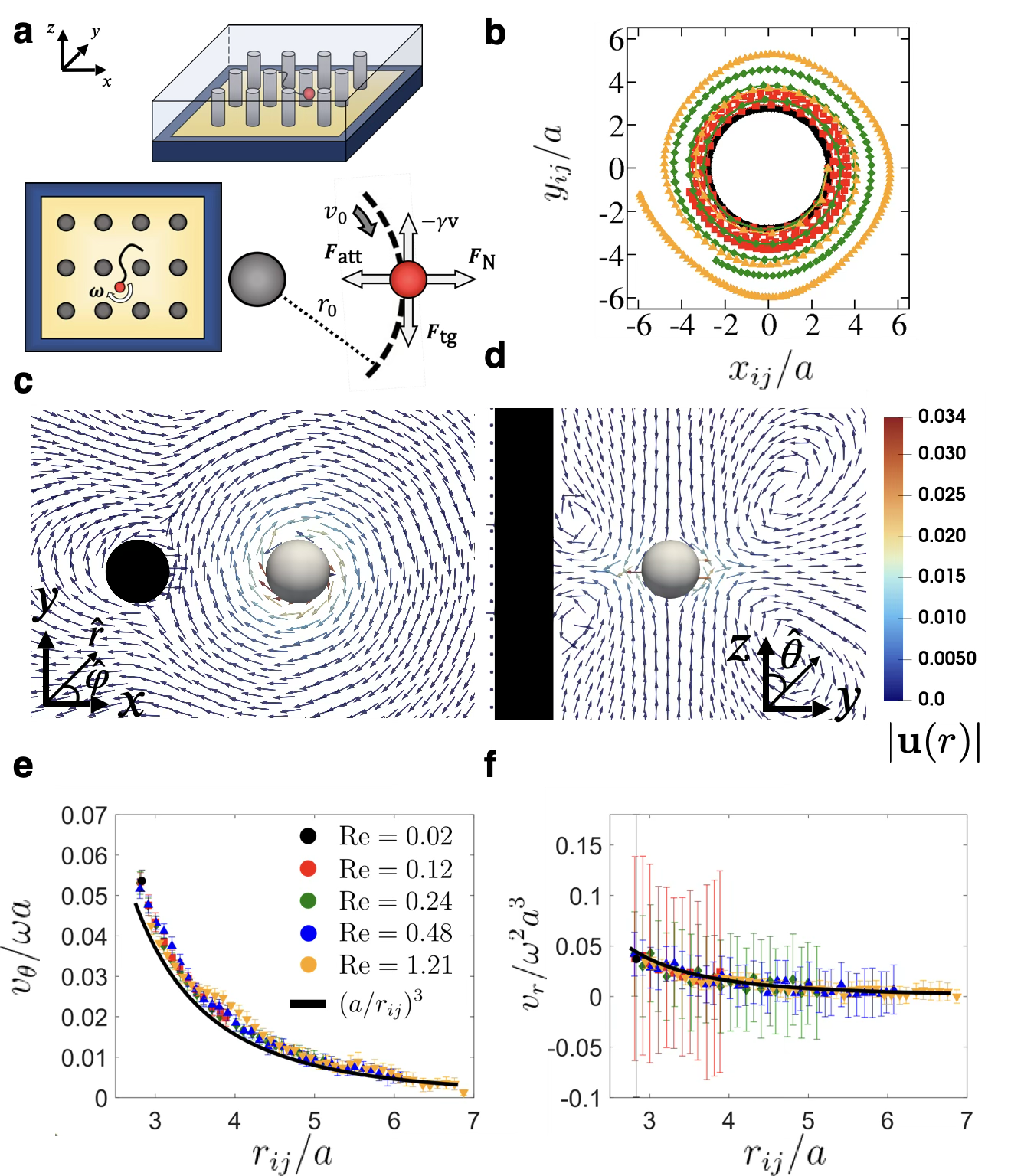}
\caption{a) Perspective (top) and overhead (bottom) view of the system setup and scheme of the forces felt by the spinner. 
b) Trajectories of spinning particles around a solid obstacle initially separated by $r_{ij}=2.75a$ at $\text{Re}=0.02$ (black circles), $\text{Re}=0.12$ (red squares), $\text{Re}=0.24$ (green diamonds) and $\text{Re}=1.21$ (orange triangles). c, d) Fluid velocity fields on the $xy$ and $zy$-plane generated by the spinner at $\text{Re}=1.2$ in the presence of an obstacle. e,f) Tangential (left) and radial (right) components of the translational velocity as a function of the distance $r_{ij}$ between the spinner and the obstacle, along the trajectory around a solid obstacle initially separated by $r_{ij}=2.75a$ at different rotational frequencies: $\text{Re}=0.02$ (black), $\text{Re}=0.12$ (red), $\text{Re}=0.24$ (green), $\text{Re}=0.48$ (blue) and $\text{Re}=1.21$ (orange).} 
\label{fig:setup}
\end{figure}

Here, we study the transport of rotating colloids, {\em spinners}, in periodic obstacle arrays. We show that the competition between inertial (Magnus-like) lift and short-range attraction generates stable orbits whose radius is set by the rotation frequency, as schematically shown in Fig.~\ref{fig:setup}a. In periodic lattices, this balance produces two robust transport modes: corner states, where particles orbit individual posts, and inner states, where orbits couple across four neighboring obstacles. Slow modulation of the rotation frequency drives reversible transitions between these modes, enabling deterministic, stepwise transport across the grid using a single control parameter. To establish this mechanism, we combine fully resolved lattice–Boltzmann simulations with a symmetry-based Langevin description that captures the effective hydrodynamic and attractive interactions. The resulting theory predicts the orbital structure and transport modes, in quantitative agreement with the simulations.

\section{Hydrodynamic forces and Lattice-Boltzmann simulations} To quantify the hydrodynamic coupling between a spinner and an obstacle, we performed fully resolved three-dimensional Lattice–Boltzmann (LB) simulations~\cite{Dunweg2008}. The fluid is solved on a D3Q19 lattice~\cite{qian1992} using a multiple-relaxation-time collision operator~\cite{dhumieres2002} with $\Delta x=1$ and time step $\Delta t=1$ and setting the density of the LB fluid to $\rho_f=1$ and its dynamic viscosity to $\eta=1/6$. The simulation domain is $N_x\times N_y\times N_z=128\times 128\times 128$, periodic in $x$ and $y$ and bounded in the $z$ direction by top and bottom no-slip walls. The spinner is modeled as a sphere of radius $a=8\Delta x$ and the obstacle as a fixed cylinder of radius $a_{\rm O}=8\Delta x$ and height $N_z \Delta x$. Both objects are implemented using the standard bounce--back rule~\cite{laddLatticeBoltzmannSimulationsParticleFluid,Ding2003}, which enforces no-slip boundary conditions at their surfaces. The spinning particle is initialized at the mid-plane ($z=N_z/2$) and is free to move in 3D; however, due to the vertical symmetry of the obstacle array and the absence of external vertical forces, the trajectory remains stable within the $xy$-plane. The spinner is actuated by a time-dependent magnetic torque $\mathbf{M} \times \mathbf{B}(t)$ that imposes a prescribed angular velocity $\omega$. This torque is applied via localized magnetic forces $\mathbf{F}_{\rm mag}$ on surface flags attached to the spinner. Hydrodynamic forces and torques on the particle are computed through momentum exchange during bounce--back. The relevant control parameter is the particle Reynolds number, $\mathrm Re = \frac{\omega a^2 \rho_f}{\eta}$, which in our simulations spans $0.02 \lesssim \mathrm{Re} \lesssim 1.2$.

A spinner in an unbounded fluid generates an azimuthal rotlet flow and exhibits no net translation. In contrast, near a boundary, rotation couples to translation via hydrodynamic interactions. To characterize this coupling, we initialize the spinner at a distance $r=2.75a$ from a cylindrical obstacle and track its motion in the plane, resolved into radial ($\hat{r}$) and azimuthal ($\hat{\varphi}$) components relative to the obstacle center. Two dynamical regimes emerge. At low rotation rates ($\mathrm{Re}\le 0.02$), the spinner follows nearly circular orbits around the post, as shown by the black trajectory in \hyperref[fig:setup]{Fig. 1b}. At higher rotation rates ($\mathrm{Re}>0.02$), the polar component , $\hat{\theta}$, of the flow~\cite{Aragones2016}, the meridional circulation in the vertical plane, generates an inertial lift that drives a slow radial migration, producing a spiral trajectory (green line in \hyperref[fig:setup]{Fig. 1b}). The tangential velocity of the spinner, $v_{\theta}$, originates from azimuthal flow scattered by the obstacle~\cite{blake1974}, which produces a secondary flow acting back on the spinner (\hyperref[fig:setup]{Fig. 1c}). This coupling decays as $v_{\theta} \sim 1/r^3$ (\hyperref[fig:setup]{Fig. 1e}). At finite Reynolds number, $\mathrm{Re} > 0.02$, the polar component of the flow also produces a radial velocity $v_r$ directed along $\bm\omega\times\mathbf v_{\theta}$ (\hyperref[fig:setup]{Fig. 1d})~\cite{Aragones2016}, generating the repulsive lift. Both $v_{\theta}$ and $v_r$ decay as $1/r^3$, and the lift eventually vanishes at large separations, as shown in \hyperref[fig:setup]{Fig. 1f}.\\

\section{Langevin description of the spinner} 
To describe the long-time dynamics, we coarse-grain the hydrodynamic interaction with a fixed post into effective forces entering a Langevin equation. A rotating sphere generates an azimuthal rotlet flow $\mathbf{v}_f(\mathbf{r})=\left(\frac{a}{r}\right)^3\boldsymbol{\omega}\times\mathbf{r}$, which, after scattering from the obstacle, produces a tangential hydrodynamic force on the spinner,
\begin{equation}
\mathbf{F}_{\text{tg}} = -\beta \frac{a^4}{r^3}\boldsymbol{\omega}\times\hat{\mathbf{r}} = \boldsymbol{\nabla}\times\mathbf{c}(\mathbf{r}), \quad
\mathbf{c}(\mathbf{r}) = -\beta a^4\frac{\omega}{2r^2}\hat{\mathbf{z}},
\label{eq:f_tg}
\end{equation}
where $\beta$ is the coupling coefficient extracted from LB simulations and $\hat{\mathbf{r}}$ is the unit vector from spinner to obstacle, $\mathbf{c}(\mathbf{r})$ is the vector potential, and $\hat{\mathbf{z}}$ is perpendicular to the substrate. At finite Reynolds number, the spinner also experiences a Magnus-like lift force, $\mathbf{F}_{\text{N}}=M_0\boldsymbol{\omega}\times\mathbf{v}$, with $M_0\sim\pi a^3\rho_f$~\cite{Rubinow1961} and $\boldsymbol{\omega}=\omega\hat{\mathbf{z}}$, which drives radial migration away from the obstacle. Thermal fluctuations are included through a stochastic force $\boldsymbol{\xi}(t)$ with zero mean and variance $\langle\xi_i(t)\xi_j(t')\rangle = 2k_B T\gamma_s \delta_{ij} \delta(t-t')$, where $k_B T$ is the thermal energy and $\gamma_s = 6\pi\eta a$ is the Stokes drag coefficient of the spinner. The coarse-grained dynamics thus obey,
\begin{equation}
    m\frac{d \mathbf v}{dt}=\mathbf F_{\text{tg}}(\mathbf r)-(\gamma_s \mathbf v-M_0\bm\omega\times\mathbf v)+ \boldsymbol{\xi}(t)
\label{eq:langevin}
\end{equation}

Since both viscous and Magnus forces depend linearly on velocity, they combine into an effective friction matrix, $\hat{\gamma}$. In terms of the Pauli matrix $\sigma_y$, it takes the compact form $\hat{\gamma}=\gamma_s(\mathbf{I}+i\nu\sigma_y)=\hat{\mu}^{-1}$ where $\nu=\frac{\omega M_0}{\gamma_s}\approx\frac{\omega a^2\rho_f}{6\eta}=\frac{\text{Re}}{6}$ quantifies the relative strength of Magnus to viscous forces. In steady state, the noise-averaged velocity is $\langle \mathbf v_j\rangle = \hat{\mu}_{jk}\langle \mathbf F_k\rangle$, and its azimuthal and radial components satisfy the scaling shown by the LB simulations,
\begin{equation}
    \frac{\langle v_{\theta}\rangle}{\omega a}=\frac{\beta/\gamma_s}{1+\nu^2}\left(\frac{a}{r}\right)^3\qquad\frac{\langle v_r\rangle}{\omega^2a^3}=\frac{\rho_f}{6\eta}\frac{\beta/\gamma_s}{1+\nu^2}\left(\frac{a}{r}\right)^3
\label{eq:velocities}
\end{equation}
From our LB simulations in the low-Re regime ($\nu < 1$), we find $\beta \approx \gamma_s$, giving $v_\theta/(\omega a) \approx (a/r)^3$ and $v_r/(\omega^2 a^3) \approx (\rho_f/6\eta)(a/r)^3$, as shown in \hyperref[fig:setup]{Fig. 1e-f}. Therefore, the ratio between the radial and tangential velocities scales as $\frac{\langle v_r\rangle}{\langle v_{\theta}\rangle} =\mathrm{Re}/6$. Since the radial velocity is suppressed by the factor $\nu$ at low Reynolds numbers, fluctuations that do not follow this scaling produce relative errors amplified by $\nu^{-1}$. We observe this in our LB simulations, where numerical noise results in larger relative error bars in the radial velocity at low $\mathrm{Re}$. This indicates that the susceptibility to other sources of noise, such as thermal fluctuations, will be enhanced in the radial direction and increased with the Reynolds number.

\section{Trapping the spinner to the obstacle}
Hydrodynamic lift pushes the spinner away from the post, but this repulsion can be counteracted by an attractive short–range interaction. For simplicity, we consider a point charge $Q$ at the obstacle center that attracts the spinner with charge $q$. Defining $F_a=\frac{Qq}{4\pi\varepsilon_0 a^2}$, the attractive force and its potential are,
\begin{equation}
    \mathbf F_{\text{att}}=-F_a\left(\frac{a}{r}\right)^2 \hat{\mathbf r}=-\bm\nabla u(\mathbf r)\qquad u(\mathbf r)=-F_a \frac{a^2}{r}
\label{eq:f_att}
\end{equation}

Balancing this attractive force with the hydrodynamic lift stabilizes the spinner on a circular orbit around the post (Fig.~\ref{fig:setup}a). The orbital behavior is governed only by the ratio between the attractive and viscous force, and therefore does not depend on the microscopic origin of the attraction. The long-time statistics of this bound state follow from the Fokker–Planck equation governing the probability density $\rho(\mathbf r,t)$,
\begin{equation}
\frac{\partial \rho(\mathbf{r}, t)}{\partial t} = 
D_{\text{eff}}\bm\nabla^2\rho
- \nabla \cdot \left( \mathbf{W} \rho \right)
= -\nabla \cdot \mathbf{j}
\label{eq:fp}
\end{equation}
where $D_{\text{eff}}=\frac{D}{1+\nu^2}$ is the effective diffusion coefficient arising from the combined viscous and Magnus contributions to the mobility tensor $\hat{\mu} = \hat{\gamma}^{-1}$. The drift velocity $\mathbf W=\hat{\mu}(\mathbf F_{\text{tg}}+\mathbf F_{\text{att}})$ combines the hydrodynamic tangential coupling and the attraction, and the conserved current is $\mathbf j=-D_{\text{eff}}\bm\nabla\rho+\mathbf W\rho$. Because $D_{\rm eff}$ decreases as $\nu$ increases, the stationary probability distribution becomes narrower at higher driving frequencies, confining the spinner more tightly to its orbit.

\begin{figure}
\includegraphics[width=1\linewidth]{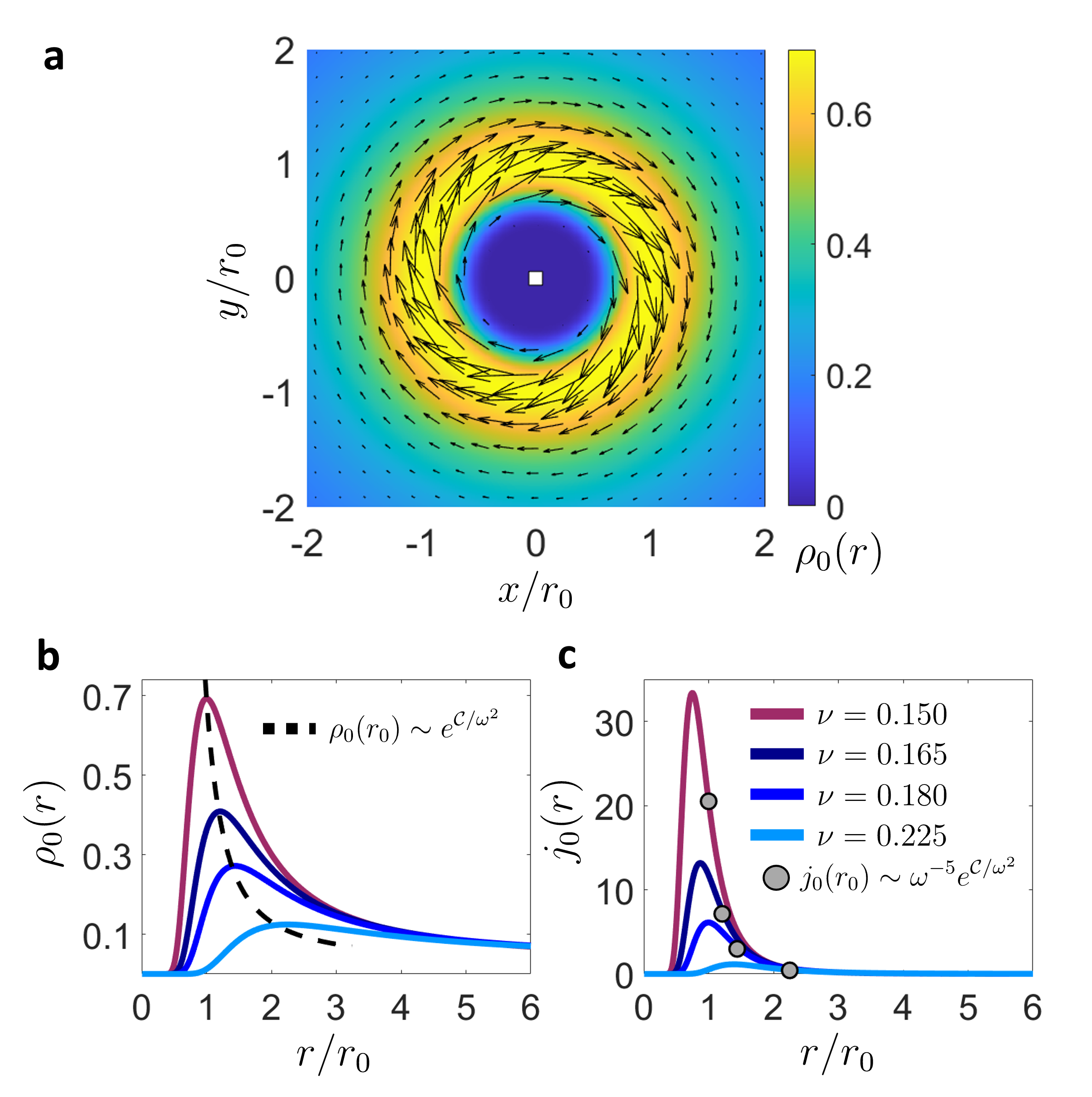}
\caption{a) Normalized probability density and current density in the $xy$ plane for the equilibrium solution, where the obstacle is located in the origin. 
b) Radial profile of stationary probability density. The dashed line corresponds to the stationary density at $r_0$ as a function of $\omega$. 
Distance values are normalized by $r_0$ evaluated at $\nu=0.150$. c) Radial profile of the stationary current. The grey dots corresponds to the stationary current at the maximum density $\mathbf j_0(r_0)$ for different $\omega$.
Distance values are normalized by $r_0$ evaluated at $\nu=0.150$.}
\label{fig:oneobs}
\end{figure}

The stationary solution follows from a Helmholtz decomposition of the drift velocity,
\begin{equation}
\begin{aligned}
&\mathbf W=-\bm\nabla U(\mathbf r)+\bm\nabla\times\mathbf A(\mathbf r) \\
U=&\frac{u-\nu \mathrm c}{\gamma_s(1+\nu^2)}\qquad \mathbf A=\frac{\mathrm c+\nu u}{\gamma_s(1+\nu^2)}\hat{\mathbf z}
\end{aligned}
\label{eq:helmholtz}
\end{equation}
where $u(\mathbf r)$ and $\mathbf c=\mathrm c(\mathbf r)\hat{\mathbf z}$ are the Coulomb potential and hydrodynamic vector potentials defined by Eq.~(\ref{eq:f_tg}) and Eq.~(\ref{eq:f_att}). For a single obstacle, the stationary density and current take the exact form,
\begin{equation}
    \rho_0(r,\omega)=\frac{1}{V_{\text{N}}}e^{-\frac{U(r)}{D_{\text{eff}}}}=\frac{1}{V_{\text{N}}}\text{exp}\left[\frac{2 E_{\text{B}}}{k_{\text{B}}T}\left(\frac{r_0}{r}-\frac{r_0^2}{2r^2}\right)\right]
\label{eq:density_sol}
\end{equation}
\begin{equation}
    \mathbf j_0(r,\omega)=-(\bm\nabla\times\mathbf A)\rho_0(r, \omega)=-\left(\frac{r_0}{r}\right)^3\frac{1+\frac{r}{r_0}\nu^2}{1+\nu^2}v_0\rho_0\hat{\bm\theta}
\label{eq:current_sol}
\end{equation}
where $V_{\text{N}}$ normalizes the density, $\hat{\bm\theta}$ is the tangential unit vector. The characteristic orbital radius $r_0$, orbital velocity $v_0$, and binding energy $E_{\text B}$ are,
\begin{equation}
    r_0=\nu\frac{\beta\omega a}{F_a}a\hspace{0.45cm} v_0=\frac{\beta}{\gamma_s}\left(\frac{a}{r_0}\right)^3\omega a\hspace{0.45cm} E_{\text{B}}=\frac{a}{r_0}\frac{F_a a}{2}
\label{eq:scales}
\end{equation}
The orbital radius grows quadratically with the rotational frequency because it is set by the balance of two different time scales of the system, $r_0 = \omega^2 \tau_1 \tau_2 a$ where $\tau_1 = M_0 / \gamma_s$ and $\tau_2 = \beta a / F_a$, associated with inertial lift and attraction. Their ratio defines a dimensionless control parameter, $\mathrm{Ga} = \tau_1/\tau_2 \propto F_a \rho / \eta^2$, which compares attractive and viscous forces and governs the orbital regime.

The maximum probability density lies at $r_0$ (\hyperref[fig:oneobs]{Fig. 2a}), with radial fluctuations of width $\Delta r \sim r_0\sqrt{k_{\mathrm B}T/E_{\mathrm B}}$. Because approaching the post is energetically costly, the density decays rapidly as $\rho_0 \sim e^{-\mathcal C/r^2}$, avoiding lubrication-dominated configurations. The corresponding current field $\mathbf j_0$ describes circular motion around the obstacle with velocity $v_0$, as shown in \hyperref[fig:oneobs]{Fig. 2c}, and at the most probable radius satisfies $\mathbf j_0(r_0)=-v_0\rho_0\hat{\bm\theta}$. More generally, the the current can be written as $\mathbf j_0(r)=-v_{\text{eff}}(r)\rho_0\hat{\bm\theta}$, where $v_{\text{eff}}(r)=\langle v_{\theta}\rangle(1+\frac{r}{r_0}\nu^2)$ combines the hydrodynamic tangential velocity Eq.~(\ref{eq:velocities}) with a correction that decays as $1/r^2$. This shift explains why the maximum current is reached at a radius slightly smaller than $r_0$ (\hyperref[fig:oneobs]{Fig. 2c}). Notably, the stationary density retains an explicit dependence on the Stokes drag $\gamma_s$, reflecting the intrinsically dissipative character of the spinner dynamics. The spinner orbital trajectory is highly controllable by frequency modulation since both $r_0$ and $v_0$ depend on rotational frequency; $r_0$ grows as $\omega^2$, while the orbital velocity decays sharply as $v_0 \sim \omega^{-5}$. At sufficiently high frequencies, the spinner becomes effectively insensitive to a single obstacle, motivating the introduction of periodic obstacle arrays to achieve robust transport control.

\section{Obstacle Array System}
\begin{figure}[t!]\label{fig:screening}
    \centering
    \includegraphics[width=1\linewidth]{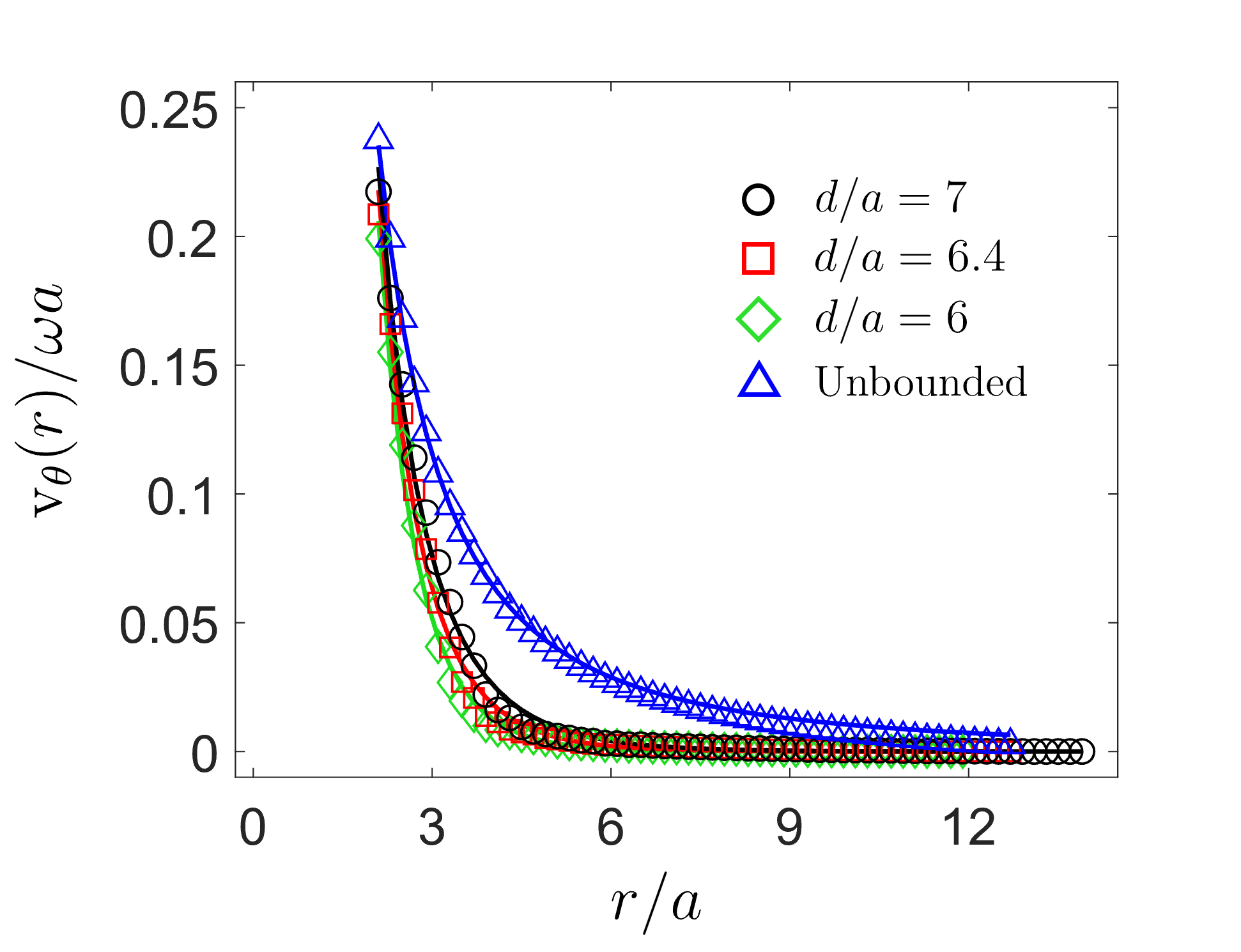}
    \caption{
    Azimuthal velocity $\mathrm{v}_\theta(r)$ generated by a sphere rotating about the $z$-axis in a square obstacle lattice with lattice spacings $d/a = 6$ (green), $6.4$ (red), and $7$ (blue). Distance is normalized by the particle radius $a$ and velocities by $\omega a$. Solid lines show fit to the screened-rotlet form $\mathrm{v}_{\theta}(r) \approx r^{-2} (1+\frac{r}{\lambda}) e^{-r/\lambda}$ with screening lengths $\lambda/a \approx 0.96$, $\lambda/a \approx 1.17$ and $\lambda/a \approx 1.47$, respectively. The unbounded rotlet results (blue triangles) follow the prediction $\mathrm{v}_\theta \sim r^{-2}$ (blue solid line).  is recovered at short distances (blue triangles), while at larger $r$ the lattice induces an exponential cutoff.}
\end{figure}

In a periodic array of obstacles the flow generated by a rotating particle experiences repeated scattering events. As a result, the long-ranged rotlet field becomes effectively screened, and its azimuthal velocity is well described by the Brinkman fluid which decays as $\mathrm{v}_{\theta}(r) \approx r^{-2} (1+\frac{r}{\lambda}) e^{-r/\lambda}$ with a lattice-dependent screening length $\lambda$ measured numerically (\hyperref[fig:screening]{Fig. 3}) ~\cite{cortez2010}. This behavior indicates that the obstacle array acts as an effective porous medium that attenuates momentum transport over distances larger than a lattice-dependent hydrodynamic screening length. Since the measured screening lengths are significantly smaller than the lattice spacing ($\lambda << d$), obstacle-mediated interactions are dominated by the exponential attenuation of the flow field. In the coarse-grained theory we therefore retain only the exponential cutoff $e^{-r/\lambda}$, which captures the finite interaction range induced by hydrodynamic screening, while neglecting the algebraic prefactors of the Brinkman solution. Motivated by this observation, we incorporate screening directly into the obstacle-mediated interactions by multiplying both the scalar $U$ and vector $\mathbf A$ potentials associated with each obstacle by the same exponential factor. The cutoff length is chosen to be of the order of the lattice spacing, $d$, ensuring that only the local obstacle environment contributes appreciably. The screened potentials are therefore written as,
\begin{equation}
    U_i=\frac{u_i-\nu \mathrm c_i}{\gamma_s(1+\nu^2)}e^{-\frac{|\mathbf r-\mathbf r_i|}{d}}\quad\mathbf A_i=\frac{\mathrm c_i+\nu u_i}{\gamma_s(1+\nu^2)}e^{-\frac{|\mathbf r-\mathbf r_i|}{d}}\hat{\mathbf z}
\end{equation}
where $r_i$ is the position of the obstacle $i$. Superposing these screened contributions yields approximate expressions for the stationary density and current, 
\begin{equation}
\rho_0(\mathbf r, \omega)=\frac{1}{V_{\text{N}}}\prod_{i=1}^{N_{\text{obs}}}\exp\left[-\frac{U_i}{D_{\text{eff}}}\right]
\label{eq:density_varobs}
\end{equation}
\begin{equation}
\mathbf j_0(\mathbf r, \omega)=-\sum_{i=1}^{N_{\text{obs}}}(\bm\nabla\times\mathbf A_i)\rho_0
\label{eq:current_varobs}
\end{equation}
which remain in excellent agreement with numerical solutions because the two driving fields, $-\bm\nabla U$ and $\bm\nabla\times\mathbf A$, remain nearly orthogonal after superposition, so the mechanism that generates a non-equilibrium steady state persists. 

\begin{figure}[t!]\label{fig:varobs}
\includegraphics[width=0.97\linewidth]{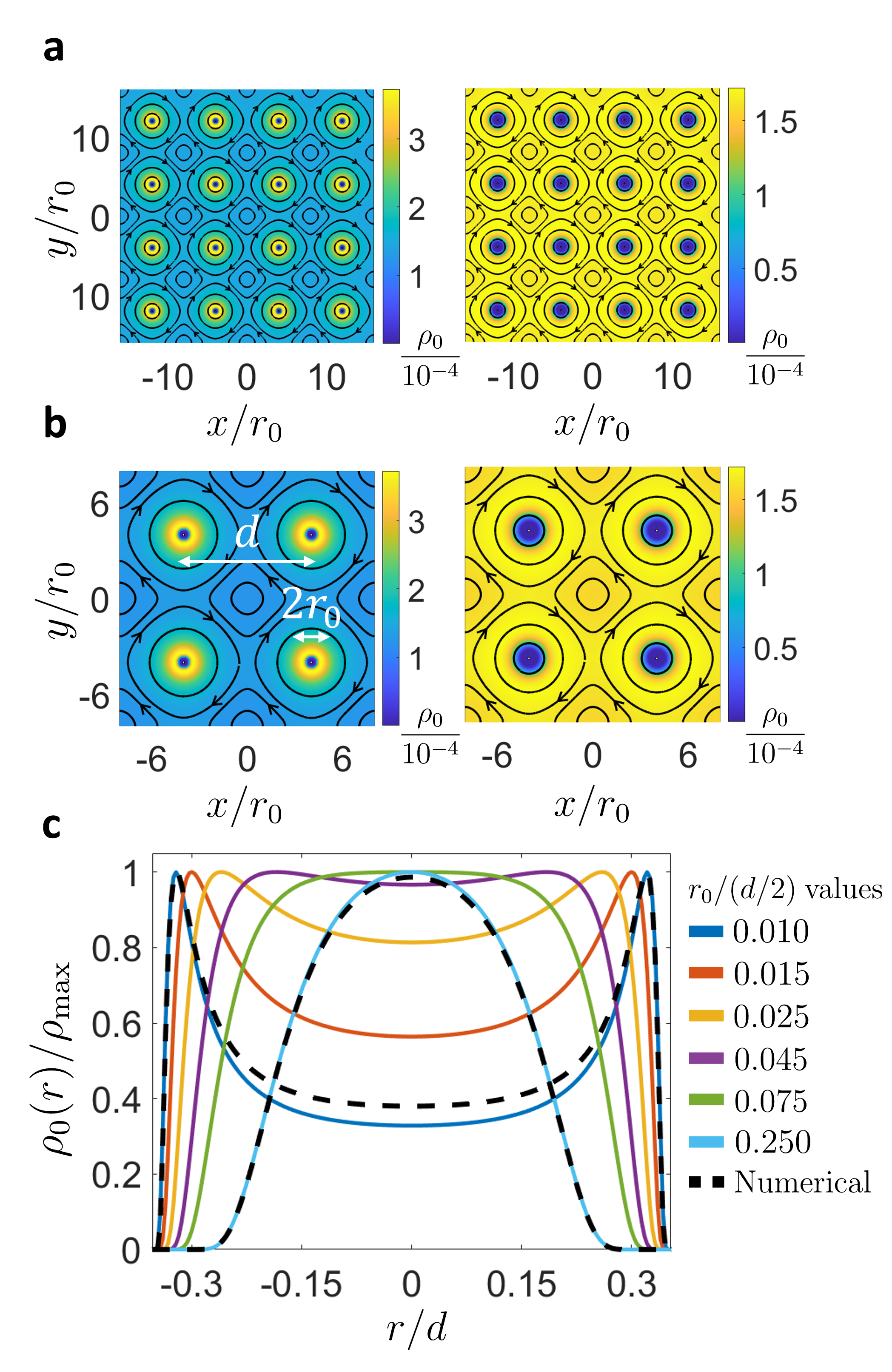}
\caption{
a) Normalized probability density in the $xy$ plane obtained from the numerical solution for Eq.(\ref{eq:fp}) for a spinner rotating at $\nu = 2.0$ (left) and $\nu = 4.0$ (right) in a 8x8 square obstacle lattice (4x4 shown). Increasing $\nu$ enhances the hydrodynamic repulsion and shifts probability from single–post orbits to the interstitial regions between four neighboring obstacles. The streamlines of the normalized probability current are superimposed. Clockwise circulations around each obstacle correspond to \emph{corner states}, whereas counterclockwise orbits in the channels between obstacles correspond to \emph{inner states}.
b) Normalized probability density in the $xy$ plane for the approximate equilibrium solution Eq.~(\ref{eq:density_varobs}) in an 8x8 obstacle grid (2x2 shown), using a spinner's rotation frequency of $\nu=2.0$ (left) and $\nu=4.0$ (right), and the streamlines showing the normalized current density given by the approximate solution Eq.~(\ref{eq:current_varobs}). There is high agreement with numerical solutions.
c) Density profiles along the diagonal of the lattices unit cell at $\nu=2.5$, which shows the crossover between corner states, found when $r_0$ is low compared to $d/2$, and inner states, found at higher $r_0/(d/2)$ values. The black dashed lines corresponds to the numerical solutions. 
}
\end{figure}

Using an 8x8 square obstacle lattice, we find the stationary density $\rho_0$ and current streamlines $\mathbf j_0$ for two representative rotational frequencies, $\nu=2.0$ and $\nu=4.0$, through numerical resolution of Eq.~(\ref{eq:fp}) in \hyperref[fig:varobs]{Fig. 4a}, and evaluating the solutions given by Eq.~(\ref{eq:density_varobs}) and Eq.(\ref{eq:current_varobs}) in \hyperref[fig:varobs]{Fig. 4b}. High agreement is observed between both, which sustains our approximation. At low frequency, Coulomb attraction dominates and the spinner interacts essentially with a single post. The density consists of disconnected circular orbits, {\em corner states}, localized around each obstacle, and the corresponding probability current displays clockwise circulation around each post (\hyperref[fig:varobs]{Fig. 4a,vb}). As the rotation frequency increases, hydrodynamic repulsion becomes significant and the orbital radius $r_0$ approaches to $d/2$. Probability accumulates in the region between four neighboring obstacles, while remaining lower at the center of the unit cell, as it can be seen in \hyperref[fig:varobs]{Fig. 4c}. In this {\em inner state}, the spinner forms counter-rotating orbits between obstacles, although corner states and weak-current diffusion through the central region still coexist. Notice that the streamlines are independent of $\nu$ (see Appendix \ref{app:A}). At still higher frequencies, repulsion becomes sufficiently strong that the spinner is pushed toward the center of each unit cell, as evidenced by the increased central density in \hyperref[fig:varobs]{Fig. 4c}. Furthermore, comparison between the analytical and numerical profiles in \hyperref[fig:varobs]{Fig. 4c} demonstrate that the screened-superposition theory given by Eqs.~\ref{eq:density_varobs}-\ref{eq:current_varobs} accurately captures the stationary density and current throughout this crossover.

\section{Dynamic control of corner and inner transport modes}
\begin{figure}
\includegraphics[width=1\linewidth]{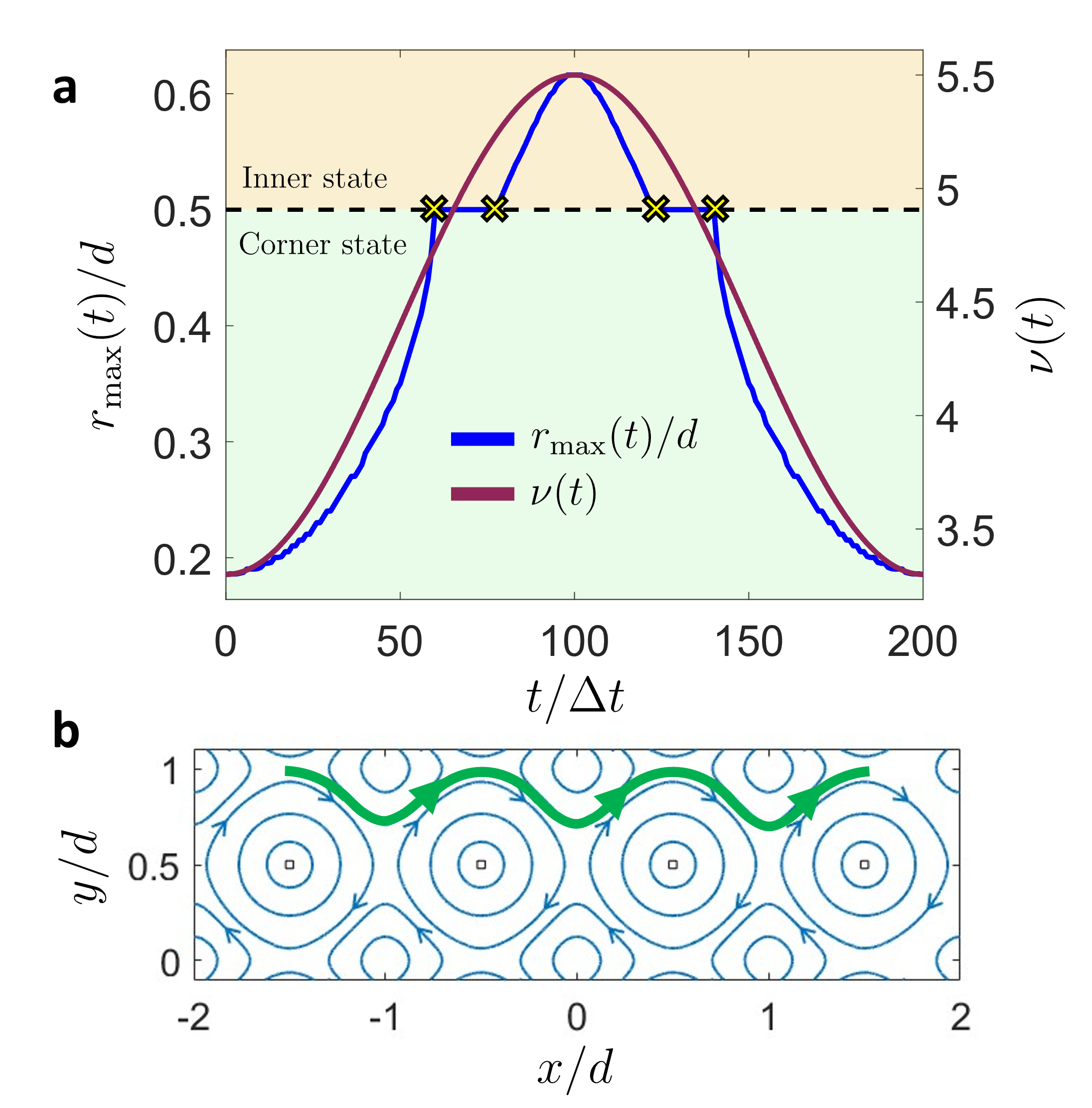}
\caption{ a) One period of a sinusoidal frequency modulation (purple) and the radius corresponding to maximum probability density as a function of time (blue). b) Schematic representation of the trajectory of the particle density due to the currents.  
}
\label{fig:transport}
\end{figure}

\begin{center}
\begin{figure}[t!]
\includegraphics[width=1\linewidth]{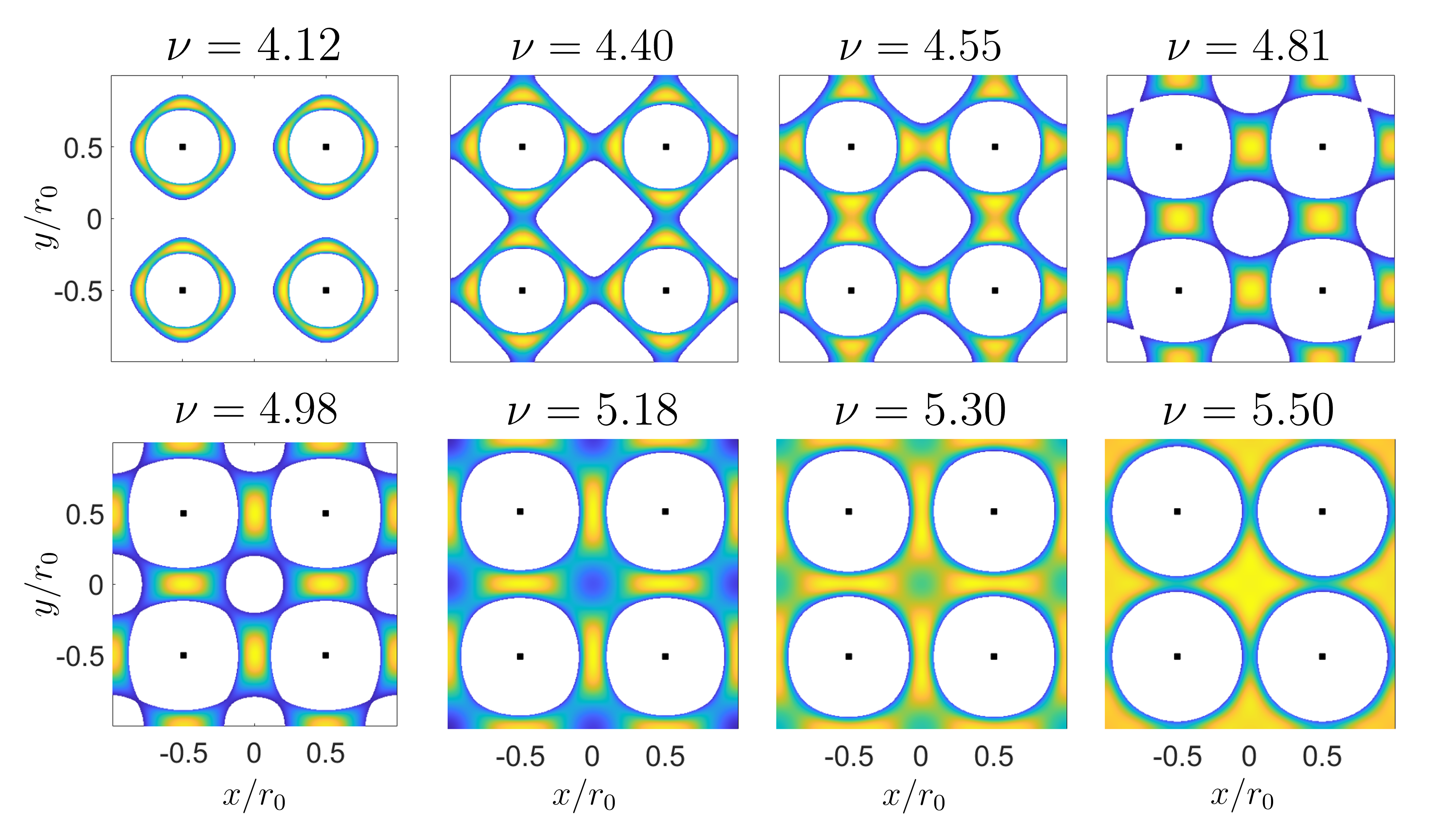}
\caption{ Evolution of high–density regions across the corner--inner crossover.
    Spatial regions where $\rho_0(\mathbf r;\nu) > 0.99\,\rho_{\max}$ for several rotational frequencies $\nu$ (increasing from left to right).  
    At low $\nu$, four isolated lobes appear around each obstacle (corner states).  
    As $\nu$ increases, these lobes merge along the midpoints between obstacles, producing a frequency window where the maximal–density point remains fixed at $r_{\max}=d/2$.  
    Upon further increase, the high–density region expands toward the centers of the four–post motifs, signaling the emergence of the inner state.  
    This structural reorganization underlies the dynamical gateway enabling programmable transport.
    }
    \label{fig:S_modulation}
\end{figure}
\end{center}

The spinners motion can be controlled dynamically by modulating its rotation frequency, allowing deterministic switching between corner and inner states. To demonstrate this mechanism, we track the instantaneous radius of maximum probability density, $r_{\max}(t)$, while driving the spinner with a time-dependent frequency $\nu(t)=\nu_0\left[1+A_0\cos\left(2\pi\frac{t}{\tau_m}\right)\right]$, where parameters $\nu_0$, $A_0$ and $\tau_m$ are chosen so that the drive sweeps quasistatically across the corner–inner crossover. As shown in \hyperref[fig:transport]{Fig. 5a} and \hyperref[fig:transport]{Fig. 5b}, $r_{\max}(t)$ follows the imposed modulation, but exhibits a plateau at $r_{\max} = d/2$, which originates from the structure of the stationary density. In the crossover regime, the analytic theory predicts the emergence of the inner state, in which orbits around adjacent posts hybridize to form a four-lobed trajectory running along the channels between obstacles. This inner state displays high density along the four sides of the unit cell, a robust counter-clockwise circulating current, and a depleted central region. Crucially, because the inner orbit is geometrically pinned by the obstacle arrangement, its radius remains fixed at $d/2$ over a finite interval of frequencies, as can be seen in \hyperref[fig:S_modulation]{Fig. 6} . The spinner remains in this pinned inner orbit until hydrodynamic repulsion becomes strong enough to push it toward the central state at higher $\nu$.

To clarify the origin of the dynamical switching mechanism, we analyze the spatial structure of the high–density regions obtained from the approximate stationary solution $\rho_0(\mathbf r;\nu)$ as the rotational frequency $\nu$ is varied. Figure~\ref{fig:S_modulation} shows, for a $2\times 2$ portion of the full lattice, the region where the density exceeds $0.99\,\rho_{\max}$ for several values of $\nu$ spanning the corner--inner crossover. At low frequencies, the attraction to each post dominates and the spinner localizes in circular orbits centered on individual obstacles. These are the corner states; the high–density region appears as four disconnected lobes, each wrapped around a single post. As $\nu$ increases, hydrodynamic repulsion becomes comparable to attraction and the orbits associated with neighboring posts begin to overlap. In this intermediate regime the geometry of the high–density lobes reorganizes: instead of expanding outward, the four corner orbits progressively merge along the midpoints of the edges connecting adjacent posts. Over the frequency interval $\nu\simeq 4.5$--$5.25$, this reorganization produces a striking effect: the position of the density maximum remains locked to the midpoint between two neighboring obstacles. The measured radius of maximum density stays fixed at $r_{\max}=d/2$, generating the plateau shown in \hyperref[fig:transport]{Fig.~5a}.

\begin{figure}[h!]
    \centering
    \includegraphics[width=0.8\linewidth]{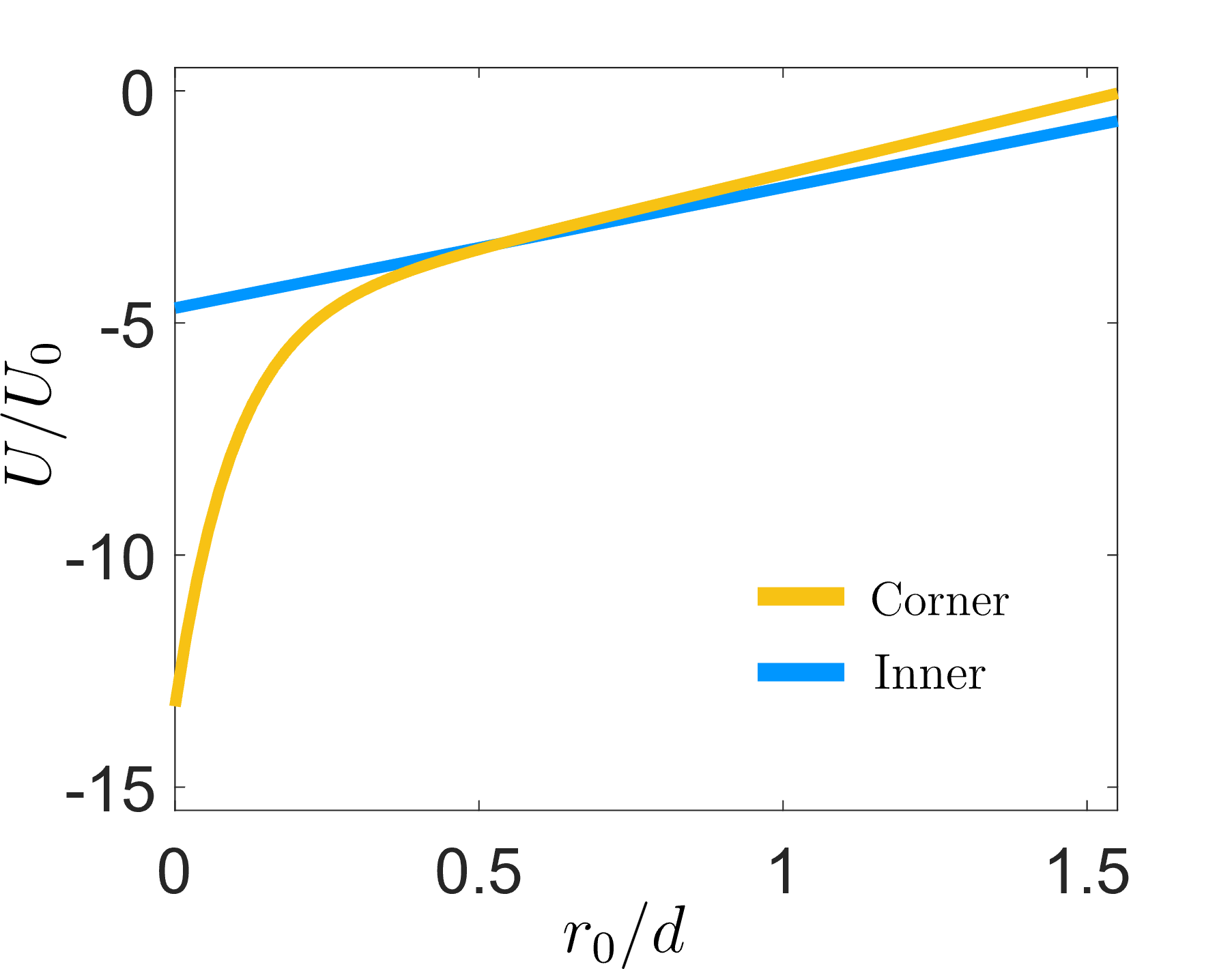}
    \caption{
    Potential energy across the corner--inner crossover.
    Orange and blue lines correspond to the potential energy of the corner and inner states, respectively, as function of the orbital radius. 
    }
    \label{fig:S_potential}
\end{figure}

This plateau arises because, in this frequency window, the most probable orbit is a geometrically constrained channel connecting neighboring posts. Increasing $\nu$ does not shift the density peak radially; instead, it broadens the corresponding lobe. The potential within the corner states is minimum until it reaches the position in between obstacles; however the potential corresponding to the inner states is still larger and thus, the particle density remains at that position until the repulsive potential overcomes the attractive one, as shown in \hyperref[fig:S_potential]{Fig.~7}. Once the broadened lobe feels the central region of the four–post motif, a lower–energy configuration becomes available: the spinner reorganizes into the inner state, characterized by probability maxima at the center of each four–obstacle cell and counter–rotating currents. Beyond this point, $r_{\max}$ begins to grow again, approaching the geometric value $d/\sqrt{2}$.

This sequence of reorganizations constitutes the  dynamical gateway: slow modulation of $\nu(t)$ reliably drives the system into the plateau state $r_{\max}=d/2$, where the orbit is pinned by lattice geometry, and from which the spinner can be steered either into a corner state or into an inner state depending on the direction of frequency change. This geometric locking is what enables robust, stepwise transport in the frequency–modulated protocol.

Specifically, when the drive increases through the crossover,  the spinner reliably hops into the inner state and begins counter-rotating around the interstitial region. When the modulation reverses, the spinner leaves the inner orbit and returns to the $d/2$ orbit, from which it can fall back into a corner state around a different obstacle. Completing a full modulation cycle can return the spinner to its original state, but the transport becomes far more robust when the drive operates predominantly in the intermediate-frequency regime where the plateau exists. In this regime, shown in \hyperref[fig:transport]{Fig. 5b}, the modulation repeatedly brings the spinner into the $r=d/2$ orbit, thereby avoiding low-frequency corner trapping and executing deterministic hops between adjacent inner orbits. Because the streamline topology is independent of $\nu$ (see Appendix \ref{app:A}), the transport direction and magnitude can be programmed directly through the temporal shape of $\nu(t)$. Thus, the obstacle lattice acts as a hydrodynamic mode selector, while the rotation frequency serves as a single control parameter that switches the particle between topologically distinct dynamical states, clockwise corner modes and counter-clockwise inner modes. By exploiting both the robustness of the inner-orbit plateau and the opposite chirality of the two states, we obtain a simple, stable, and broadly applicable protocol for actively transporting microrotors across structured environments.\\

\section{Conclusions}
We have shown that a single rotating colloid navigating a structured environment can be steered between distinct hydrodynamic transport modes simply by tuning its rotation frequency. Hydrodynamic scattering, inertial lift, and a short-range attraction combine to generate robust corner and inner orbits, whose geometry and chirality are accurately captured by a coarse-grained Fokker–Planck description informed by fully resolved simulations. In periodic lattices, these modes reorganize into spatially extended steady states whose occupation can be controlled dynamically. Slow modulation of the driving frequency induces reversible transitions between the two modes and creates a finite interval where trajectories lock to the lattice geometry, enabling deterministic stepwise transport across the grid. Our results establish a minimal and predictive mechanism, controlled by the ratio between attractive and viscous forces, for programming the motion of active rotors in complex environments and open a route toward steering dense suspensions of interacting spinners using hydrodynamic design alone.
\onecolumngrid

\twocolumngrid

\begin{acknowledgments}
\noindent The authors acknowledge financial support by MCIN/AEI/10.13039/501100011033/ for all grants listed next: JVA for PID2022-139776NB-C64 supported by "ERDF A way of making Europe"; JLA for PID2022-143010NB-I00 and CEX2023-001316-M, also supported by "ERDF A way of making Europe"; JLA for RYC2019-028189-I and CNS2023-145447, also supported by "European Union NextGenerationEU/PRTR". AAK is grateful to MISTI Spain for funding and to the Michael and Sonja Koerner chair for financial support as well.
\end{acknowledgments}

\newpage

\appendix
\section{Invariance of the streamlines with respect to $\omega$}
\label{app:A}
As shown, density current for a grid of obstacles can be written as:
\begin{equation}
\begin{aligned}
\mathbf j_0(\mathbf r, &\omega)=-\sum_{i}(\bm\nabla\times\mathbf A_i)\rho_0 \\
\mathbf j_0=-\frac{\rho_0}{\gamma(1+\nu^2)}&\sum_i\bm\nabla\times\left[(\mathrm c_i+\nu u_i)e^{-\frac{|\mathbf r-\mathbf r_i|}{d}}\hat{\mathbf z}\right]
\end{aligned}
\label{eq:apA_current}
\end{equation}

Notice we have included the cutoff previously discussed. To calculate $\mathbf j_0$, we have to take the curl of both terms:
\begin{equation}
\begin{aligned}
&\bm\nabla\times\left[(\mathrm c_i+\nu u_i)e^{-\frac{|\mathbf r-\mathbf r_i|}{d}}\hat{\mathbf z}\right]=\\
=e^{-\frac{|\mathbf r-\mathbf r_i|}{d}}\bm\nabla\times&\left[(\mathrm c_i+\nu u_i)\hat{\mathbf z}\right]+(\mathrm c_i+\nu u_i)\bm\nabla\times\left(e^{-\frac{|\mathbf r-\mathbf r_i|}{d}}\hat{\mathbf z}\right)
\end{aligned}
\label{eq:apA_curl}
\end{equation}
Computing both curls gives the following result:
\begin{equation}
\begin{aligned}
&\mathbf j_0=-\frac{v_0\rho_0r_0^3}{1+\nu^2}\sum_ig_i(|\mathbf r-\mathbf r_i|)\hat{\bm\theta}_i\\
g_i(|\mathbf r-\mathbf r_i|)=&\frac{1}{|\mathbf r-\mathbf r_i|^2}\left(\frac{1+h_i}{|\mathbf r-\mathbf r_i|}+\frac{1+2 h_i}{2d}\right)e^{-\frac{|\mathbf r-\mathbf r_i|}{d}}\\
\end{aligned}
\label{eq:apA_solution}
\end{equation}

We have defined $h_i=\frac{|\mathbf r-\mathbf r_i|}{r_0}\nu^2$ and  $\hat{\bm\theta}_i=\frac{(-(y-y_i),x-x_i)}{|\mathbf r-\mathbf r_i|}$. Notice that the streamlines' direction is determined by $\sum_i g_i\hat{\bm\theta}_i$, and not by the factor in front which indeed depends on $\omega$ and only contributes to the current's modulus. In principle, function $g_i$ could depend on $\omega$ through quantities $\nu^2$ and $r_0$. However, $\nu^2\sim\omega^2$ and $r_0\sim\omega^2$, so the quotient $\nu^2/r_0$ is independent of the angular frequency and so does $g_i$ and the current's direction. This justifies choosing an intermediate frequency in \hyperref[fig:transport]{Fig. 5b} to represent the streamlines. It should be said that the invariance of the streamlines with respect to the angular frequency is a property independent of the specific cutoff function used, as long as it does not depend on $\omega$.


\begin{thebibliography}{32}%
\makeatletter
\providecommand \@ifxundefined [1]{%
 \@ifx{#1\undefined}
}%
\providecommand \@ifnum [1]{%
 \ifnum #1\expandafter \@firstoftwo
 \else \expandafter \@secondoftwo
 \fi
}%
\providecommand \@ifx [1]{%
 \ifx #1\expandafter \@firstoftwo
 \else \expandafter \@secondoftwo
 \fi
}%
\providecommand \natexlab [1]{#1}%
\providecommand \enquote  [1]{``#1''}%
\providecommand \bibnamefont  [1]{#1}%
\providecommand \bibfnamefont [1]{#1}%
\providecommand \citenamefont [1]{#1}%
\providecommand \href@noop [0]{\@secondoftwo}%
\providecommand \href [0]{\begingroup \@sanitize@url \@href}%
\providecommand \@href[1]{\@@startlink{#1}\@@href}%
\providecommand \@@href[1]{\endgroup#1\@@endlink}%
\providecommand \@sanitize@url [0]{\catcode `\\12\catcode `\$12\catcode
  `\&12\catcode `\#12\catcode `\^12\catcode `\_12\catcode `\%12\relax}%
\providecommand \@@startlink[1]{}%
\providecommand \@@endlink[0]{}%
\providecommand \url  [0]{\begingroup\@sanitize@url \@url }%
\providecommand \@url [1]{\endgroup\@href {#1}{\urlprefix }}%
\providecommand \urlprefix  [0]{URL }%
\providecommand \Eprint [0]{\href }%
\providecommand \doibase [0]{https://doi.org/}%
\providecommand \selectlanguage [0]{\@gobble}%
\providecommand \bibinfo  [0]{\@secondoftwo}%
\providecommand \bibfield  [0]{\@secondoftwo}%
\providecommand \translation [1]{[#1]}%
\providecommand \BibitemOpen [0]{}%
\providecommand \bibitemStop [0]{}%
\providecommand \bibitemNoStop [0]{.\EOS\space}%
\providecommand \EOS [0]{\spacefactor3000\relax}%
\providecommand \BibitemShut  [1]{\csname bibitem#1\endcsname}%
\let\auto@bib@innerbib\@empty
\bibitem [{\citenamefont {Berg}\ and\ \citenamefont
  {Purcell}(1977)}]{berg1977}%
  \BibitemOpen
  \bibfield  {author} {\bibinfo {author} {\bibfnamefont {H.}~\bibnamefont
  {Berg}}\ and\ \bibinfo {author} {\bibfnamefont {E.}~\bibnamefont {Purcell}},\
  }\bibfield  {title} {\bibinfo {title} {Physics of chemoreception},\ }\href
  {https://doi.org/10.1016/S0006-3495(77)85544-6} {\bibfield  {journal}
  {\bibinfo  {journal} {Biophysical Journal}\ }\textbf {\bibinfo {volume}
  {20}},\ \bibinfo {pages} {193} (\bibinfo {year} {1977})}\BibitemShut
  {NoStop}%
\bibitem [{\citenamefont {Fortunato}\ \emph {et~al.}(2024)\citenamefont
  {Fortunato}, \citenamefont {Br{\"u}ckner}, \citenamefont {Grosser},
  \citenamefont {Rossetti}, \citenamefont {{Bosch-Padr{\'o}s}}, \citenamefont
  {Trebicka}, \citenamefont {{Roca-Cusachs}}, \citenamefont {Sunyer},
  \citenamefont {Hannezo},\ and\ \citenamefont {Trepat}}]{fortunato2024}%
  \BibitemOpen
  \bibfield  {author} {\bibinfo {author} {\bibfnamefont {I.~C.}\ \bibnamefont
  {Fortunato}}, \bibinfo {author} {\bibfnamefont {D.~B.}\ \bibnamefont
  {Br{\"u}ckner}}, \bibinfo {author} {\bibfnamefont {S.}~\bibnamefont
  {Grosser}}, \bibinfo {author} {\bibfnamefont {L.}~\bibnamefont {Rossetti}},
  \bibinfo {author} {\bibfnamefont {M.}~\bibnamefont {{Bosch-Padr{\'o}s}}},
  \bibinfo {author} {\bibfnamefont {J.}~\bibnamefont {Trebicka}}, \bibinfo
  {author} {\bibfnamefont {P.}~\bibnamefont {{Roca-Cusachs}}}, \bibinfo
  {author} {\bibfnamefont {R.}~\bibnamefont {Sunyer}}, \bibinfo {author}
  {\bibfnamefont {E.}~\bibnamefont {Hannezo}},\ and\ \bibinfo {author}
  {\bibfnamefont {X.}~\bibnamefont {Trepat}},\ }\href
  {https://doi.org/10.1101/2024.12.02.626413} {\bibinfo {title} {Single cell
  migration along and against confined haptotactic gradients}} (\bibinfo {year}
  {2024})\BibitemShut {NoStop}%
\bibitem [{\citenamefont {Mierke}(2020)}]{mierke2020}%
  \BibitemOpen
  \bibfield  {author} {\bibinfo {author} {\bibfnamefont {C.~T.}\ \bibnamefont
  {Mierke}},\ }\bibfield  {title} {\bibinfo {title} {Mechanical {{Cues Affect
  Migration}} and {{Invasion}} of {{Cells From Three Different Directions}}},\
  }\href {https://doi.org/10.3389/fcell.2020.583226} {\bibfield  {journal}
  {\bibinfo  {journal} {Frontiers in Cell and Developmental Biology}\ }\textbf
  {\bibinfo {volume} {8}},\ \bibinfo {pages} {583226} (\bibinfo {year}
  {2020})}\BibitemShut {NoStop}%
\bibitem [{\citenamefont {Lozano}\ \emph {et~al.}(2016)\citenamefont {Lozano},
  \citenamefont {Ten~Hagen}, \citenamefont {L{\"o}wen},\ and\ \citenamefont
  {Bechinger}}]{lozano2016}%
  \BibitemOpen
  \bibfield  {author} {\bibinfo {author} {\bibfnamefont {C.}~\bibnamefont
  {Lozano}}, \bibinfo {author} {\bibfnamefont {B.}~\bibnamefont {Ten~Hagen}},
  \bibinfo {author} {\bibfnamefont {H.}~\bibnamefont {L{\"o}wen}},\ and\
  \bibinfo {author} {\bibfnamefont {C.}~\bibnamefont {Bechinger}},\ }\bibfield
  {title} {\bibinfo {title} {Phototaxis of synthetic microswimmers in optical
  landscapes},\ }\href {https://doi.org/10.1038/ncomms12828} {\bibfield
  {journal} {\bibinfo  {journal} {Nature Communications}\ }\textbf {\bibinfo
  {volume} {7}},\ \bibinfo {pages} {12828} (\bibinfo {year}
  {2016})}\BibitemShut {NoStop}%
\bibitem [{\citenamefont {Koley}\ and\ \citenamefont {Nanda}()}]{koley}%
  \BibitemOpen
  \bibfield  {author} {\bibinfo {author} {\bibfnamefont {S.}~\bibnamefont
  {Koley}}\ and\ \bibinfo {author} {\bibfnamefont {K.~K.}\ \bibnamefont
  {Nanda}},\ }\bibfield  {title} {\bibinfo {title} {Algae-like {{Artificial
  Organic Photo-tactic Micro-swimmers}}},\ }\href@noop {} {\ }\BibitemShut
  {NoStop}%
\bibitem [{\citenamefont {Wan}\ \emph {et~al.}(2008)\citenamefont {Wan},
  \citenamefont {Olson~Reichhardt}, \citenamefont {Nussinov},\ and\
  \citenamefont {Reichhardt}}]{wanRectificationSwimmingBacteria2008}%
  \BibitemOpen
  \bibfield  {author} {\bibinfo {author} {\bibfnamefont {M.~B.}\ \bibnamefont
  {Wan}}, \bibinfo {author} {\bibfnamefont {C.~J.}\ \bibnamefont
  {Olson~Reichhardt}}, \bibinfo {author} {\bibfnamefont {Z.}~\bibnamefont
  {Nussinov}},\ and\ \bibinfo {author} {\bibfnamefont {C.}~\bibnamefont
  {Reichhardt}},\ }\bibfield  {title} {\bibinfo {title} {Rectification of
  {{Swimming Bacteria}} and {{Self-Driven Particle Systems}} by {{Arrays}} of
  {{Asymmetric Barriers}}},\ }\href
  {https://doi.org/10.1103/PhysRevLett.101.018102} {\bibfield  {journal}
  {\bibinfo  {journal} {Physical Review Letters}\ }\textbf {\bibinfo {volume}
  {101}},\ \bibinfo {pages} {018102} (\bibinfo {year} {2008})}\BibitemShut
  {NoStop}%
\bibitem [{\citenamefont {Nikola}\ \emph {et~al.}(2016)\citenamefont {Nikola},
  \citenamefont {Solon}, \citenamefont {Kafri}, \citenamefont {Kardar},
  \citenamefont {Tailleur},\ and\ \citenamefont
  {Voituriez}}]{nikolaActiveParticlesSoft2016}%
  \BibitemOpen
  \bibfield  {author} {\bibinfo {author} {\bibfnamefont {N.}~\bibnamefont
  {Nikola}}, \bibinfo {author} {\bibfnamefont {A.~P.}\ \bibnamefont {Solon}},
  \bibinfo {author} {\bibfnamefont {Y.}~\bibnamefont {Kafri}}, \bibinfo
  {author} {\bibfnamefont {M.}~\bibnamefont {Kardar}}, \bibinfo {author}
  {\bibfnamefont {J.}~\bibnamefont {Tailleur}},\ and\ \bibinfo {author}
  {\bibfnamefont {R.}~\bibnamefont {Voituriez}},\ }\bibfield  {title} {\bibinfo
  {title} {Active {{Particles}} with {{Soft}} and {{Curved Walls}}:
  {{Equation}} of {{State}}, {{Ratchets}}, and {{Instabilities}}},\ }\href
  {https://doi.org/10.1103/PhysRevLett.117.098001} {\bibfield  {journal}
  {\bibinfo  {journal} {Physical Review Letters}\ }\textbf {\bibinfo {volume}
  {117}},\ \bibinfo {pages} {098001} (\bibinfo {year} {2016})}\BibitemShut
  {NoStop}%
\bibitem [{\citenamefont {Lauga}(2016)}]{Lauga2016}%
  \BibitemOpen
  \bibfield  {author} {\bibinfo {author} {\bibfnamefont {E.}~\bibnamefont
  {Lauga}},\ }\bibfield  {title} {\bibinfo {title} {Bacterial
  {{Hydrodynamics}}},\ }\href
  {https://doi.org/10.1146/annurev-fluid-122414-034606} {\bibfield  {journal}
  {\bibinfo  {journal} {Annual Review of Fluid Mechanics}\ }\textbf {\bibinfo
  {volume} {48}},\ \bibinfo {pages} {105} (\bibinfo {year} {2016})},\ \Eprint
  {https://arxiv.org/abs/1509.02184} {arXiv:1509.02184} \BibitemShut {NoStop}%
\bibitem [{\citenamefont {{Brun-Cosme-Bruny}}\ \emph
  {et~al.}(2020)\citenamefont {{Brun-Cosme-Bruny}}, \citenamefont
  {F{\"o}rtsch}, \citenamefont {Zimmermann}, \citenamefont {Bertin},
  \citenamefont {Peyla},\ and\ \citenamefont
  {Rafa{\"i}}}]{brun-cosme-brunyDeflectionPhototacticMicroswimmers2020}%
  \BibitemOpen
  \bibfield  {author} {\bibinfo {author} {\bibfnamefont {M.}~\bibnamefont
  {{Brun-Cosme-Bruny}}}, \bibinfo {author} {\bibfnamefont {A.}~\bibnamefont
  {F{\"o}rtsch}}, \bibinfo {author} {\bibfnamefont {W.}~\bibnamefont
  {Zimmermann}}, \bibinfo {author} {\bibfnamefont {E.}~\bibnamefont {Bertin}},
  \bibinfo {author} {\bibfnamefont {P.}~\bibnamefont {Peyla}},\ and\ \bibinfo
  {author} {\bibfnamefont {S.}~\bibnamefont {Rafa{\"i}}},\ }\bibfield  {title}
  {\bibinfo {title} {Deflection of phototactic microswimmers through obstacle
  arrays},\ }\href {https://doi.org/10.1103/PhysRevFluids.5.093302} {\bibfield
  {journal} {\bibinfo  {journal} {Physical Review Fluids}\ }\textbf {\bibinfo
  {volume} {5}},\ \bibinfo {pages} {093302} (\bibinfo {year}
  {2020})}\BibitemShut {NoStop}%
\bibitem [{\citenamefont {Reichhardt}\ and\ \citenamefont
  {Reichhardt}(2020)}]{reichhardtDirectionalLockingEffects2020}%
  \BibitemOpen
  \bibfield  {author} {\bibinfo {author} {\bibfnamefont {C.}~\bibnamefont
  {Reichhardt}}\ and\ \bibinfo {author} {\bibfnamefont {C.~J.~O.}\ \bibnamefont
  {Reichhardt}},\ }\bibfield  {title} {\bibinfo {title} {Directional locking
  effects for active matter particles coupled to a periodic substrate},\ }\href
  {https://doi.org/10.1103/PhysRevE.102.042616} {\bibfield  {journal} {\bibinfo
   {journal} {Physical Review E}\ }\textbf {\bibinfo {volume} {102}},\ \bibinfo
  {pages} {042616} (\bibinfo {year} {2020})}\BibitemShut {NoStop}%
\bibitem [{\citenamefont {Spagnolie}\ \emph {et~al.}(2015)\citenamefont
  {Spagnolie}, \citenamefont {{Moreno-Flores}}, \citenamefont {Bartolo},\ and\
  \citenamefont {Lauga}}]{spagnolie2015a}%
  \BibitemOpen
  \bibfield  {author} {\bibinfo {author} {\bibfnamefont {S.~E.}\ \bibnamefont
  {Spagnolie}}, \bibinfo {author} {\bibfnamefont {G.~R.}\ \bibnamefont
  {{Moreno-Flores}}}, \bibinfo {author} {\bibfnamefont {D.}~\bibnamefont
  {Bartolo}},\ and\ \bibinfo {author} {\bibfnamefont {E.}~\bibnamefont
  {Lauga}},\ }\bibfield  {title} {\bibinfo {title} {Geometric capture and
  escape of a microswimmer colliding with an obstacle},\ }\href
  {https://doi.org/10.1039/C4SM02785J} {\bibfield  {journal} {\bibinfo
  {journal} {Soft Matter}\ }\textbf {\bibinfo {volume} {11}},\ \bibinfo {pages}
  {3396} (\bibinfo {year} {2015})}\BibitemShut {NoStop}%
\bibitem [{\citenamefont {Avron}(1998)}]{Avron1998}%
  \BibitemOpen
  \bibfield  {author} {\bibinfo {author} {\bibfnamefont {J.~E.}\ \bibnamefont
  {Avron}},\ }\bibfield  {title} {\bibinfo {title} {Odd viscosity},\ }\href {https://doi.org/10.1023/A:1023084404080} {\bibfield
  {journal} {\bibinfo  {journal} {Journal of Statistical Physics}\ }\textbf
  {\bibinfo {volume} {92}},\ \bibinfo {pages} {543} (\bibinfo {year} {1998})},\
  \Eprint {https://arxiv.org/abs/physics/9712050} {arXiv:physics/9712050}
  \BibitemShut {NoStop}%
\bibitem [{\citenamefont {Banerjee}\ \emph {et~al.}(2017)\citenamefont
  {Banerjee}, \citenamefont {Souslov}, \citenamefont {Abanov},\ and\
  \citenamefont {Vitelli}}]{Banerjee2017}%
  \BibitemOpen
  \bibfield  {author} {\bibinfo {author} {\bibfnamefont {D.}~\bibnamefont
  {Banerjee}}, \bibinfo {author} {\bibfnamefont {A.}~\bibnamefont {Souslov}},
  \bibinfo {author} {\bibfnamefont {A.~G.}\ \bibnamefont {Abanov}},\ and\
  \bibinfo {author} {\bibfnamefont {V.}~\bibnamefont {Vitelli}},\ }\bibfield
  {title} {\bibinfo {title} {Odd viscosity in chiral active fluids},\ }\href
  {https://doi.org/10.1038/s41467-017-01378-7} {\bibfield  {journal} {\bibinfo
  {journal} {Nature Communications}\ }\textbf {\bibinfo {volume} {8}},\
  \bibinfo {pages} {1} (\bibinfo {year} {2017})},\ \Eprint
  {https://arxiv.org/abs/1702.02393} {arXiv:1702.02393} \BibitemShut {NoStop}%
\bibitem [{\citenamefont {Grzybowski}\ and\ \citenamefont
  {Whitesides}(2002)}]{Grzybowski2002}%
  \BibitemOpen
  \bibfield  {author} {\bibinfo {author} {\bibfnamefont {B.~A.}\ \bibnamefont
  {Grzybowski}}\ and\ \bibinfo {author} {\bibfnamefont {G.~M.}\ \bibnamefont
  {Whitesides}},\ }\bibfield  {title} {\bibinfo {title} {Dynamic aggregation of
  chiral spinners},\ }\href {https://doi.org/10.1126/science.1068130}
  {\bibfield  {journal} {\bibinfo  {journal} {Science}\ }\textbf {\bibinfo
  {volume} {296}},\ \bibinfo {pages} {718} (\bibinfo {year}
  {2002})}\BibitemShut {NoStop}%
\bibitem [{\citenamefont {G{\"o}tze}\ and\ \citenamefont
  {Gompper}(2011)}]{Gotze2011}%
  \BibitemOpen
  \bibfield  {author} {\bibinfo {author} {\bibfnamefont {I.~O.}\ \bibnamefont
  {G{\"o}tze}}\ and\ \bibinfo {author} {\bibfnamefont {G.}~\bibnamefont
  {Gompper}},\ }\bibfield  {title} {\bibinfo {title} {Dynamic self-assembly and
  directed flow of rotating colloids in microchannels},\ }\href
  {https://doi.org/10.1103/PhysRevE.84.031404} {\bibfield  {journal} {\bibinfo
  {journal} {Physical Review E - Statistical, Nonlinear, and Soft Matter
  Physics}\ }\textbf {\bibinfo {volume} {84}},\ \bibinfo {pages} {031404}
  (\bibinfo {year} {2011})}\BibitemShut {NoStop}%
\bibitem [{\citenamefont {Climent}\ \emph {et~al.}(2007)\citenamefont
  {Climent}, \citenamefont {Yeo}, \citenamefont {Maxey},\ and\ \citenamefont
  {Karniadakis}}]{climentDynamicSelfAssemblySpinning2007}%
  \BibitemOpen
  \bibfield  {author} {\bibinfo {author} {\bibfnamefont {E.}~\bibnamefont
  {Climent}}, \bibinfo {author} {\bibfnamefont {K.}~\bibnamefont {Yeo}},
  \bibinfo {author} {\bibfnamefont {M.~R.}\ \bibnamefont {Maxey}},\ and\
  \bibinfo {author} {\bibfnamefont {G.~E.}\ \bibnamefont {Karniadakis}},\
  }\bibfield  {title} {\bibinfo {title} {Dynamic {{Self-Assembly}} of
  {{Spinning Particles}}},\ }\href {https://doi.org/10.1115/1.2436587}
  {\bibfield  {journal} {\bibinfo  {journal} {Journal of Fluids Engineering}\
  }\textbf {\bibinfo {volume} {129}},\ \bibinfo {pages} {379} (\bibinfo {year}
  {2007})}\BibitemShut {NoStop}%
\bibitem [{\citenamefont {Goto}\ and\ \citenamefont {Tanaka}(2015)}]{Goto2015}%
  \BibitemOpen
  \bibfield  {author} {\bibinfo {author} {\bibfnamefont {Y.}~\bibnamefont
  {Goto}}\ and\ \bibinfo {author} {\bibfnamefont {H.}~\bibnamefont {Tanaka}},\
  }\bibfield  {title} {\bibinfo {title} {Purely hydrodynamic ordering of
  rotating disks at a finite {{Reynolds}} number},\ }\href
  {https://doi.org/10.1038/ncomms6994} {\bibfield  {journal} {\bibinfo
  {journal} {Nature Communications}\ }\textbf {\bibinfo {volume} {6}},\
  \bibinfo {pages} {5994} (\bibinfo {year} {2015})}\BibitemShut {NoStop}%
\bibitem [{\citenamefont {Aragones}\ \emph {et~al.}(2019)\citenamefont
  {Aragones}, \citenamefont {Steimel},\ and\ \citenamefont
  {{Alexander-Katz}}}]{Aragones2019}%
  \BibitemOpen
  \bibfield  {author} {\bibinfo {author} {\bibfnamefont {J.~L.}\ \bibnamefont
  {Aragones}}, \bibinfo {author} {\bibfnamefont {P.}~\bibnamefont {Steimel}},\
  and\ \bibinfo {author} {\bibfnamefont {A.}~\bibnamefont {{Alexander-katz}}},\
  }\bibfield  {title} {\bibinfo {title} {Aggregation dynamics of active
  rotating particles in dense passive media},\ }\bibfield  {journal} {\bibinfo
  {journal} {Soft Matter}\ }\href {https://doi.org/10.1039/c8sm02207k}
  {10.1039/c8sm02207k} (\bibinfo {year} {2019})\BibitemShut {NoStop}%
\bibitem [{\citenamefont {Fily}\ \emph {et~al.}(2012)\citenamefont {Fily},
  \citenamefont {Baskaran},\ and\ \citenamefont {Marchetti}}]{fily2012b}%
  \BibitemOpen
  \bibfield  {author} {\bibinfo {author} {\bibfnamefont {Y.}~\bibnamefont
  {Fily}}, \bibinfo {author} {\bibfnamefont {A.}~\bibnamefont {Baskaran}},\
  and\ \bibinfo {author} {\bibfnamefont {M.~C.}\ \bibnamefont {Marchetti}},\
  }\bibfield  {title} {\bibinfo {title} {Cooperative self-propulsion of active
  and passive rotors},\ }\href {https://doi.org/10.1039/c2sm06952k} {\bibfield
  {journal} {\bibinfo  {journal} {Soft Matter}\ }\textbf {\bibinfo {volume}
  {8}},\ \bibinfo {pages} {3002} (\bibinfo {year} {2012})}\BibitemShut
  {NoStop}%
\bibitem [{\citenamefont {Gorce}\ \emph {et~al.}(2021)\citenamefont {Gorce},
  \citenamefont {Bliokh}, \citenamefont {Xia}, \citenamefont {Francois},
  \citenamefont {Punzmann},\ and\ \citenamefont
  {Shats}}]{gorceRollingSpinnersWater2021}%
  \BibitemOpen
  \bibfield  {author} {\bibinfo {author} {\bibfnamefont {J.-B.}\ \bibnamefont
  {Gorce}}, \bibinfo {author} {\bibfnamefont {K.~Y.}\ \bibnamefont {Bliokh}},
  \bibinfo {author} {\bibfnamefont {H.}~\bibnamefont {Xia}}, \bibinfo {author}
  {\bibfnamefont {N.}~\bibnamefont {Francois}}, \bibinfo {author}
  {\bibfnamefont {H.}~\bibnamefont {Punzmann}},\ and\ \bibinfo {author}
  {\bibfnamefont {M.}~\bibnamefont {Shats}},\ }\bibfield  {title} {\bibinfo
  {title} {Rolling spinners on the water surface},\ }\href
  {https://doi.org/10.1126/sciadv.abd4632} {\bibfield  {journal} {\bibinfo
  {journal} {Science Advances}\ }\textbf {\bibinfo {volume} {7}},\ \bibinfo
  {pages} {eabd4632} (\bibinfo {year} {2021})}\BibitemShut {NoStop}%
\bibitem [{\citenamefont {Rubinow}\ and\ \citenamefont
  {Keller}(1961)}]{Rubinow1961}%
  \BibitemOpen
  \bibfield  {author} {\bibinfo {author} {\bibfnamefont {S.~I.}\ \bibnamefont
  {Rubinow}}\ and\ \bibinfo {author} {\bibfnamefont {J.~B.}\ \bibnamefont
  {Keller}},\ }\bibfield  {title} {\bibinfo {title} {The transverse force on a
  spinning sphere moving in a viscous fluid},\ }\href
  {https://doi.org/10.1017/S0022112061000640} {\bibfield  {journal} {\bibinfo
  {journal} {Journal of Fluid Mechanics}\ }\textbf {\bibinfo {volume} {11}},\
  \bibinfo {pages} {447} (\bibinfo {year} {1961})}\BibitemShut {NoStop}%
\bibitem [{\citenamefont {Saffman}(1965)}]{saffman1965}%
  \BibitemOpen
  \bibfield  {author} {\bibinfo {author} {\bibfnamefont {P.~G.}\ \bibnamefont
  {Saffman}},\ }\bibfield  {title} {\bibinfo {title} {The lift on a small
  sphere in a slow shear flow},\ }\href
  {https://doi.org/10.1017/S0022112065000824} {\bibfield  {journal} {\bibinfo
  {journal} {Journal of Fluid Mechanics}\ }\textbf {\bibinfo {volume} {22}},\
  \bibinfo {pages} {385} (\bibinfo {year} {1965})}\BibitemShut {NoStop}%
\bibitem [{\citenamefont {Aragones}\ \emph {et~al.}(2016)\citenamefont
  {Aragones}, \citenamefont {Steimel},\ and\ \citenamefont
  {{Alexander-Katz}}}]{Aragones2016}%
  \BibitemOpen
  \bibfield  {author} {\bibinfo {author} {\bibfnamefont {J.~L.}\ \bibnamefont
  {Aragones}}, \bibinfo {author} {\bibfnamefont {J.~P.}\ \bibnamefont
  {Steimel}},\ and\ \bibinfo {author} {\bibfnamefont {A.}~\bibnamefont
  {{Alexander-Katz}}},\ }\bibfield  {title} {\bibinfo {title}
  {Elasticity-induced force reversal between active spinning particles in dense
  passive media},\ }\href {https://doi.org/10.1038/ncomms11325} {\bibfield
  {journal} {\bibinfo  {journal} {Nature Communications}\ }\textbf {\bibinfo
  {volume} {7}},\ \bibinfo {pages} {11325} (\bibinfo {year}
  {2016})}\BibitemShut {NoStop}%
\bibitem [{\citenamefont {Cao}\ \emph {et~al.}(2023)\citenamefont {Cao},
  \citenamefont {Das}, \citenamefont {Windbacher}, \citenamefont {Ginot},
  \citenamefont {Kr{\"u}ger},\ and\ \citenamefont
  {Bechinger}}]{caoMemoryinducedMagnusEffect2023}%
  \BibitemOpen
  \bibfield  {author} {\bibinfo {author} {\bibfnamefont {X.}~\bibnamefont
  {Cao}}, \bibinfo {author} {\bibfnamefont {D.}~\bibnamefont {Das}}, \bibinfo
  {author} {\bibfnamefont {N.}~\bibnamefont {Windbacher}}, \bibinfo {author}
  {\bibfnamefont {F.}~\bibnamefont {Ginot}}, \bibinfo {author} {\bibfnamefont
  {M.}~\bibnamefont {Kr{\"u}ger}},\ and\ \bibinfo {author} {\bibfnamefont
  {C.}~\bibnamefont {Bechinger}},\ }\bibfield  {title} {\bibinfo {title}
  {Memory-induced {{Magnus}} effect},\ }\href
  {https://doi.org/10.1038/s41567-023-02213-1} {\bibfield  {journal} {\bibinfo
  {journal} {Nature Physics}\ }\textbf {\bibinfo {volume} {19}},\ \bibinfo
  {pages} {1904} (\bibinfo {year} {2023})}\BibitemShut {NoStop}%
\bibitem [{\citenamefont {Yazdi}\ \emph {et~al.}(2020)\citenamefont {Yazdi},
  \citenamefont {Aragones}, \citenamefont {Coulter},\ and\ \citenamefont
  {{Alexander-Katz}}}]{Yazdi2020}%
  \BibitemOpen
  \bibfield  {author} {\bibinfo {author} {\bibfnamefont {S.}~\bibnamefont
  {Yazdi}}, \bibinfo {author} {\bibfnamefont {J.~L.}\ \bibnamefont {Aragones}},
  \bibinfo {author} {\bibfnamefont {J.}~\bibnamefont {Coulter}},\ and\ \bibinfo
  {author} {\bibfnamefont {A.}~\bibnamefont {{Alexander-Katz}}},\ }\bibfield
  {title} {\bibinfo {title} {Metamaterials for {{Active Colloid Transport}}},\
  }\href@noop {} {\ ,\ \bibinfo {pages} {1} (\bibinfo {year} {2020})},\ \Eprint
  {https://arxiv.org/abs/2002.06477} {arXiv:2002.06477} \BibitemShut {NoStop}%
\bibitem [{\citenamefont {D{\"u}nweg}\ and\ \citenamefont
  {Ladd}(2008)}]{Dunweg2008}%
  \BibitemOpen
  \bibfield  {author} {\bibinfo {author} {\bibfnamefont {B.}~\bibnamefont
  {D{\"u}nweg}}\ and\ \bibinfo {author} {\bibfnamefont {A.~J.~C.}\ \bibnamefont
  {Ladd}},\ }\bibfield  {title} {\bibinfo {title} {Lattice {{Boltzmann
  Simulations}} of {{Soft Matter Systems}}},\ }\href
  {https://doi.org/10.1007/12} {\bibfield  {journal} {\bibinfo  {journal}
  {Advances in Polymer Science}\ \bibinfo {pages} {221, 89}} (\bibinfo {year}
  {2008})},\ \Eprint {https://arxiv.org/abs/0808.2157} {arXiv:0808.2157}
  \BibitemShut {NoStop}%
\bibitem [{\citenamefont {Qian}\ \emph {et~al.}(1992)\citenamefont {Qian},
  \citenamefont {D'Humi{\`e}res},\ and\ \citenamefont {Lallemand}}]{qian1992}%
  \BibitemOpen
  \bibfield  {author} {\bibinfo {author} {\bibfnamefont {Y.~H.}\ \bibnamefont
  {Qian}}, \bibinfo {author} {\bibfnamefont {D.}~\bibnamefont
  {D'Humi{\`e}res}},\ and\ \bibinfo {author} {\bibfnamefont {P.}~\bibnamefont
  {Lallemand}},\ }\bibfield  {title} {\bibinfo {title} {Lattice {{BGK Models}}
  for {{Navier-Stokes Equation}}},\ }\href
  {https://doi.org/10.1209/0295-5075/17/6/001} {\bibfield  {journal} {\bibinfo
  {journal} {Europhysics Letters (EPL)}\ }\textbf {\bibinfo {volume} {17}},\
  \bibinfo {pages} {479} (\bibinfo {year} {1992})}\BibitemShut {NoStop}%
\bibitem [{\citenamefont {{d'Humi{\`e}res}}(2002)}]{dhumieres2002}%
  \BibitemOpen
  \bibfield  {author} {\bibinfo {author} {\bibfnamefont {D.}~\bibnamefont
  {{d'Humi{\`e}res}}},\ }\bibfield  {title} {\bibinfo {title}
  {Multiple--relaxation--time lattice {{Boltzmann}} models in three
  dimensions},\ }\href {https://doi.org/10.1098/rsta.2001.0955} {\bibfield
  {journal} {\bibinfo  {journal} {Philosophical Transactions of the Royal
  Society of London. Series A: Mathematical, Physical and Engineering
  Sciences}\ }\textbf {\bibinfo {volume} {360}},\ \bibinfo {pages} {437}
  (\bibinfo {year} {2002})}\BibitemShut {NoStop}%
\bibitem [{\citenamefont {Ladd}\ and\ \citenamefont
  {Verberg}(2001)}]{laddLatticeBoltzmannSimulationsParticleFluid}%
  \BibitemOpen
  \bibfield  {author} {\bibinfo {author} {\bibfnamefont {A.~J.~C.}\
  \bibnamefont {Ladd}}\ and\ \bibinfo {author} {\bibfnamefont {R.}~\bibnamefont
  {Verberg}},\ }\bibfield  {title} {\bibinfo {title} {Lattice-{{Boltzmann
  Simulations}} of {{Particle-Fluid Suspensions}}},\ }\href@noop {} {\bibfield
  {journal} {\bibinfo  {journal} {Journal of Statistical Physics}\ }\textbf
  {\bibinfo {volume} {104}},\ \bibinfo {pages} {1191} (\bibinfo {year}
  {2001})}\BibitemShut {NoStop}%
\bibitem [{\citenamefont {Ding}\ and\ \citenamefont {Aidun}(2003)}]{Ding2003}%
  \BibitemOpen
  \bibfield  {author} {\bibinfo {author} {\bibfnamefont {E.~J.}\ \bibnamefont
  {Ding}}\ and\ \bibinfo {author} {\bibfnamefont {C.~K.}\ \bibnamefont
  {Aidun}},\ }\bibfield  {title} {\bibinfo {title} {Extension of the
  {{Lattice-Boltzmann Method}} for {{Direct Simulation}} of {{Suspended
  Particles Near Contact}}},\ }\href {https://doi.org/10.1023/A:1023880126272}
  {\bibfield  {journal} {\bibinfo  {journal} {Journal of Statistical Physics}\
  }\textbf {\bibinfo {volume} {112}},\ \bibinfo {pages} {685} (\bibinfo {year}
  {2003})}\BibitemShut {NoStop}%
\bibitem [{\citenamefont {Blake}\ and\ \citenamefont
  {Chwang}(1974)}]{blake1974}%
  \BibitemOpen
  \bibfield  {author} {\bibinfo {author} {\bibfnamefont {J.~R.}\ \bibnamefont
  {Blake}}\ and\ \bibinfo {author} {\bibfnamefont {A.~T.}\ \bibnamefont
  {Chwang}},\ }\bibfield  {title} {\bibinfo {title} {Fundamental singularities
  of viscous flow: {{Part I}}: {{The}} image systems in the vicinity of a
  stationary no-slip boundary},\ }\href {https://doi.org/10.1007/BF02353701}
  {\bibfield  {journal} {\bibinfo  {journal} {Journal of Engineering
  Mathematics}\ }\textbf {\bibinfo {volume} {8}},\ \bibinfo {pages} {23}
  (\bibinfo {year} {1974})}\BibitemShut {NoStop}%
\bibitem [{\citenamefont {Cortez}\ \emph {et~al.}(2010)\citenamefont {Cortez},
  \citenamefont {Cummins}, \citenamefont {Leiderman},\ and\ \citenamefont
  {Varela}}]{cortez2010}%
  \BibitemOpen
  \bibfield  {author} {\bibinfo {author} {\bibfnamefont {R.}~\bibnamefont
  {Cortez}}, \bibinfo {author} {\bibfnamefont {B.}~\bibnamefont {Cummins}},
  \bibinfo {author} {\bibfnamefont {K.}~\bibnamefont {Leiderman}},\ and\
  \bibinfo {author} {\bibfnamefont {D.}~\bibnamefont {Varela}},\ }\bibfield
  {title} {\bibinfo {title} {Computation of three-dimensional {{Brinkman}}
  flows using regularized methods},\ }\href
  {https://doi.org/10.1016/j.jcp.2010.06.012} {\bibfield  {journal} {\bibinfo
  {journal} {Journal of Computational Physics}\ }\textbf {\bibinfo {volume}
  {229}},\ \bibinfo {pages} {7609} (\bibinfo {year} {2010})}\BibitemShut
  {NoStop}%
\end{thebibliography}
\end{document}